\documentclass{elsart} 

\begin{document}
\begin{frontmatter}
\title{Gauge invariance and massive torsionic scalar field} 
\author[IU]{Nikodem J. Pop\l awski\corauthref{cor}}
\corauth[cor]{Corresponding author}
\ead{nipoplaw@indiana.edu}
\address[IU]{Department of Physics, Indiana University, 
727 East Third Street, Bloomington, IN 47405, USA}

\begin{abstract}
The Hojman--Rosenbaum--Ryan--Shepley dynamical theory of torsion
preserves local gauge invariance of electrodynamics and makes minimal coupling
compatible with torsion. 
It also allows propagation of torsion in vacuum.
It is known that implications of this model disagree with the 
E\"{o}tv\"{o}s--Dicke--Braginsky solar tests of the principle of equivalence.
We modify this theory and make it consistent with experiment by introducing a 
massive component of the torsionic potential. 
\end{abstract}

\end{frontmatter}

Local gauge invariance is the basic principle of electrodynamics.
It is related to conservation of the electric charge and the fact that 
photons are massless~\cite{LL}.
Let us consider a local gauge transformation of a complex scalar field $\psi$:
\begin{equation}
\psi\rightarrow\psi'=e^{iq\Lambda}\psi,
\label{gauge}
\end{equation}
where $q$ is the electromagnetic coupling constant and $\Lambda$ is a function
of spacetime, $\Lambda=\Lambda(x)$.
The usual derivative of the scalar field is not gauge invariant.
Neither is the Lagrangian for this field which contains $\partial_{\mu}\psi$.
We can modify the definition of the derivative by adding to the gradient
operator a compensating field (electromagnetic potential) $A_{\mu}$:
\begin{equation}
D_{\mu}=\partial_{\mu}-iqb_{\mu}^{\,\,\,\nu}A_{\nu}.
\label{deriv}
\end{equation}
This prescription generalizes the usual procedure, for which $b_{\mu}^{\,\,\,\nu}=\delta_{\mu}^{\nu}$,
and is due to Hojman, Rosenbaum, Ryan, and Shepley (HRRS)~\cite{HRRS}.
The electromagnetic covariant derivative is gauge invariant, 
$D_{\mu}\rightarrow D_{\mu}\psi'=e^{iq\Lambda}D_{\mu}\psi$, if the 
electromagnetic potential transforms as
\begin{equation}
A_{\mu}\rightarrow A_{\mu}'=A_{\mu}+c_{\mu}^{\,\,\,\nu}\Lambda_{,\nu},
\label{pot}
\end{equation}
and the tensors $b_{\mu}^{\,\,\,\nu}$ and $c_{\mu}^{\,\,\,\nu}$ are reciprocal:
\begin{equation}
b_{\mu}^{\,\,\,\rho}c_{\rho}^{\,\,\,\nu}=\delta_{\mu}^{\nu}.
\end{equation}

In general relativity, a minimally coupled theory is constructed by replacing
the metric of special relativity $\eta_{\mu\nu}$ with the metric of general 
relativity $g_{\mu\nu}$, and by replacing ordinary derivatives with covariant 
derivatives (the comma-semicolon rule)~\cite{LL}.
The electromagnetic field tensor in curved spacetime is thus given by
\begin{equation}
F_{\mu\nu}=A_{\nu;\mu}-A_{\mu;\nu},
\label{Faraday}
\end{equation}
where $A_{\nu;\mu}=A_{\nu,\mu}-\Gamma^{\,\,\rho}_{\mu\,\nu}A_{\rho}$ is the
covariant derivative of the vector $A_{\mu}$ and the connection coefficients 
$\Gamma^{\,\,\rho}_{\mu\,\nu}$ 
are the Christoffel symbols $\{^{\,\,\rho}_{\mu\,\nu}\}$.
In the presence of torsion 
$S^{\rho}_{\,\,\,\mu\nu}=\Gamma^{\,\,\,\,\rho}_{[\mu\,\nu]}$ the connection is
\begin{equation}
\Gamma^{\,\,\rho}_{\mu\,\nu}=\{^{\,\,\rho}_{\mu\,\nu}\}+S^{\rho}_{\,\,\,\mu\nu}-2S^{\,\,\,\,\,\,\,\,\,\rho}_{(\mu\nu)},
\label{conn}
\end{equation}
which results from metric compatibility of the connection, 
$g_{\mu\nu;\rho}=g_{\mu\nu:\rho}=0$~\cite{H}.
The colon indicates the covariant differentiation with respect
to the Christoffel symbols.

The electromagnetic field tensor in a spacetime with torsion is given by 
$F_{\mu\nu}=A_{\nu,\mu}-A_{\mu,\nu}-2S^{\rho}_{\,\,\,\mu\nu}A_{\rho}$, 
and is gauge invariant provided that~\cite{HRRS}~\footnote[1]{
The HRRS theory is the only model based on the principle of minimal coupling that has an interaction between torsion and the electromagnetic field without loosing gauge invariance. It is possible to define $F_{\mu\nu}=A_{\nu,\mu}-A_{\mu,\nu}$ which is both covariant and gauge invariant for any torsion tensor, however, this prescription violates the principle of minimal coupling~\cite{H,CF}. Which way is correct should be ultimately answered by experiment.}
\begin{eqnarray}
& & b_{\mu}^{\,\,\,\nu}=e^{-\phi}\delta_{\mu}^{\nu}, \\
& & c_{\mu}^{\,\,\,\nu}=e^{\phi}\delta_{\mu}^{\nu}, \\
& & S^{\rho}_{\,\,\,\mu\nu}=\frac{1}{2}(\delta_{\nu}^{\rho}\phi_{,\mu}-\delta_{\mu}^{\rho}\phi_{,\nu}).
\label{comp}
\end{eqnarray}
The last relation means that the torsion tensor is fully determined in terms
of the torsion vector $S_{\mu}=S^{\nu}_{\,\,\,\nu\mu}$ which has a potential 
$\phi$ (torsionic scalar field or tlaplon~\cite{HRRS}).
The case $\phi=0$ corresponds to a torsionless spacetime and reproduces the 
usual form of gauge invariance.

The HRRS Lagrangian density for the gravitational field has the same form as 
that in the Einstein--Cartan theory:
\begin{equation}
\pounds_{g}=\frac{1}{16\pi}R(\Gamma)\sqrt{-g},
\label{grav}
\end{equation}
where $R(\Gamma)=R_{\mu\nu}(\Gamma)g^{\mu\nu}$ and the Ricci tensor
$R_{\mu\nu}(\Gamma)=\Gamma^{\,\,\rho}_{\nu\,\mu,\rho}-\Gamma^{\,\,\rho}_{\rho\,\mu,\nu}+\Gamma^{\,\,\sigma}_{\nu\,\mu}\Gamma^{\,\,\rho}_{\rho\,\sigma}-\Gamma^{\,\,\sigma}_{\rho\,\mu}\Gamma^{\,\,\rho}_{\nu\,\sigma}$ is derived 
from the connection $\Gamma$.
We use the units in which $c=G=\hbar=1$.
For the vector torsion with a potential~(\ref{comp}), the curvature scalar
can be decomposed into the usual torsionless Ricci scalar $R(g)$, the 
part which is solely a function of the torsionic potential, and the part
which is a total covariant divergence with respect to the Christoffel
symbols and does not contribute to the equations of field:~\cite{HRRS}
\begin{equation}
R(\Gamma)=R(g)-6\phi_{,\mu}\phi^{,\mu}+\mbox{total divergence}.
\label{decomp}
\end{equation}
We also introduce the mass of the tlaplon field $m$ for the 
reasons explained below.
In the original HRRS theory $m=0$.

The total gauge invariant action for the electromagnetic field and the 
gravitational field with torsion is given by
\begin{equation}
S=\int d^{4}x\sqrt{-g}\Bigl(-\frac{R(g)}{16\pi}+\frac{3}{8\pi}\phi_{,\mu}\phi^{,\mu}-\frac{3}{8\pi}m^{2}\phi^{2}-\frac{1}{16\pi}F_{\mu\nu}F^{\mu\nu}\Bigr).
\label{act}
\end{equation}
The equations of field are obtained from variation of $g_{\mu\nu}$, $\phi$, 
and $A^{\mu}$:
\begin{eqnarray}
& & G_{\mu\nu}=6\Bigl(\phi_{,\mu}\phi_{,\nu}-\frac{1}{2}\phi_{,\rho}\phi^{,\rho}g_{\mu\nu}\Bigr)+3m^{2}\phi^{2}g_{\mu\nu} \nonumber \\
& & +2\Bigl(\frac{1}{4}F_{\rho\sigma}F^{\rho\sigma}g_{\mu\nu}-F_{\mu\rho}F_{\nu}^{\,\,\,\rho}\Bigr), \\
& & \phi^{:\mu}_{\,\,\,\,\,\mu}+m^{2}\phi+\frac{1}{3}(F^{\mu\nu}A_{\nu})_{:\mu}=0, \\
& & F^{\mu\nu}_{\,\,\,\,\,\,\,:\nu}+F^{\mu\nu}\phi_{,\nu}=0,
\label{eof}
\end{eqnarray}
where $G_{\mu\nu}$ denotes the usual Einstein tensor.
The last two equations yield
\begin{equation}
\phi^{:\mu}_{\,\,\,\,\,\mu}+m^{2}\phi=-\frac{1}{6}F_{\mu\nu}F^{\mu\nu}.
\label{tors}
\end{equation}

We now follow~\cite{N} to examine the consistence of the modified
HRRS theory with experiment.
For the scalar field of the Sun, equation~(\ref{tors}) becomes
\begin{equation}
\nabla^{2}\phi-m^{2}\phi=\frac{1}{3}({\bf B}^{2}-{\bf E}^{2}),
\label{weak}
\end{equation}
where ${\bf B}$ and ${\bf E}$ denote respectively the magnetic and 
electric field.
The solution of equation~(\ref{weak}) outside the Sun is
\begin{equation}
\phi=\frac{2}{3}\frac{e^{-mr}}{r}E_{ne},
\label{sol}
\end{equation}
where $E_{ne}$ is the total nuclear electric energy of the Sun
and other energies are negligible.
Using the data from~\cite{N} we find
\begin{equation}
\phi=0.67\times10^{-4}U\cdot e^{-mr},
\label{num}
\end{equation}
where $r$ is the distance from the Sun and $U$ is the Newtonian
potential.
This expression modifies the relative acceleration between
gold/platinum and aluminum (see \cite{N} and references therein) by the
factor $e^{-mR}$, where $R$ is the distance of the Earth from the Sun:
\begin{equation}
{\bf a}_{rel}=2\times10^{-7}\nabla U\cdot e^{-mR}.
\label{exp}
\end{equation}

To make the presented theory compatible with experiment, we must have
\begin{equation}
2\times10^{-7}\nabla U\cdot e^{-mR}<10^{-12}\nabla U.
\label{cons}
\end{equation} 
Clearly, the case $m=0$ violates this inequality.
Equation~(\ref{cons}) gives a lower limit on the mass of the
torsionic scalar field:
\begin{equation}
m>10^{-25}GeV,
\label{res}
\end{equation}
which is way below the masses of known elementary particles.
The need for introducing a mass of a scalar field coupled to the 
electromagnetic field, to avoid violations of the principle of
equivalence in the solar system, was pointed out in~\cite{G},
and the same lower limit on this mass was found. 
There is no upper limit for $m$ so the Higgs boson would be a good
candidate for the tlaplon.
Since the lower limit on the mass of the Higgs boson is on the
order of $100GeV$, the deviations from the principle of
equivalence would be unnoticable (below $10^{-27}$ of present
experimental precision).

The conclusion of this study is that one can have a compatible
electrodynamics in the presence of torsion
without violating the principle of minimal coupling.
The HRRS prescription is a viable classical theory of coupling between
the electromagnetic field and torsion, provided that the latter
is described in terms of the scalar torsionic potential, and that
such a scalar field is massive.

\end{document}